\documentclass{aa}  

\usepackage{graphicx}
\usepackage{txfonts}
\usepackage{float}
\usepackage{ulem}
\usepackage{xcolor}

\usepackage{array}
\newcolumntype{L}{>{\centering\arraybackslash}m{2cm}}
\usepackage[colorlinks=true,linkcolor=blue,citecolor=blue, urlcolor=blue]{hyperref}

\definecolor{green2}{rgb}{.1,.6,.3}

\begin{document}

 \title{Exploring magnetic field properties at the boundary of solar pores: A comparative study based on SDO-HMI observations}

   \author{J. I. Campos Rozo
          \inst{1}
          \and
          S. Vargas Domínguez\inst{2}
          \and
          D. Utz
          \inst{3}
          \and
          A. M. Veronig\inst{1}
          \and
          A. Hanslmeier\inst{1}
          }

   \institute{IGAM, Institute of Physics, University of Graz, Universitätsplatz 5, 8010 Graz, Austria\\
              \email{jose.campos-rozo@uni-graz.at}
                      \and
             Universidad Nacional de Colombia, Observatorio Astronómico Nacional, Bogotá, Colombia
             \and
             Neuroimaging Science and Support Center-NISSC. Institute of Neurology, Kepler University Hospital, Linz, Austria.}


  \abstract
   {The Sun's magnetic fields play an important role in various solar phenomena. Solar pores are regions of intensified magnetic field strength compared to the surrounding photospheric environment, and their study can help us better understand the properties and behaviour of magnetic fields in the Sun. In this work, we investigate the properties of magnetic fields on the boundaries of solar pores, specifically focusing on the evolution of the vertical magnetic field.}
   {Up to now, there exists only a single study on magnetic field properties at the boundary region of a pore. Therefore, the main goal of this work is to increase the statistics of magnetic properties determining the pore boundary region. To this aim, we study the change of the vertical magnetic field on the boundaries of six solar pores and their time evolution.}
   {We analyse six solar pores using data from the Helioseismic and Magnetic Imager instrument on board the Solar Dynamics Observatory. We apply image processing techniques to extract the relevant features of the solar pores and determine the boundary conditions of the magnetic fields. For each pore, the maximal vertical magnetic field is determined, and the obtained results are compared with the above-mentioned previous study.}
   {We find the maximal vertical magnetic field values on the boundaries of the studied solar pores to range from 1400~G to 1600~G, with a standard deviation between 7.8\% and 14.8\%. These values are lower than those reported in the mentioned preceding study. However, this can be explained by differences in spatial resolution as well as the type of data we used. For all the pores, we find that the magnetic inclination angle lies in a range of $30\pm7$°, which is consistent with the idea that the magnetic field configuration in solar pores is mainly vertical.} 
   { The vertical magnetic field is an important factor in determining the boundary of solar pores, and it plays a more relevant role than the intensity gradient. The obtained information will be useful for future studies on the formation and evolution of magnetic structures of the Sun. Additionally, this study highlights the importance of high spatial resolution data for the purpose of accurately characterising the magnetic properties of solar pores. Overall, the findings of this work contribute to the understanding of the magnetic field properties of the Sun and will be crucial for improving models of solar dynamics and magnetic flux emergence.}

\titlerunning{Exploring magnetic field properties at the boundary of solar pores}

   \maketitle
%

\section{Introduction}
During the past few years, the study of the Sun has experienced an incredible advancement with the launch of space-based telescopes and instruments, such as the Parker Solar Probe \citep{PSP2016} and the Solar Orbiter \citep{solarorbiter2020} as well as the first light from the ground-based four-meter solar telescope the Daniel K. Inouye Solar Telescope \citep{dkist2020}, located on  Haleakalā, Maui, Hawaii. The understanding of the (small-scale) dynamics of the Sun, the interaction of the plasma and the magnetic fields, and the processes leading to large-scale eruptions in the solar atmosphere require deeper study and understanding.

The emergence of magnetic flux from the solar interior causes the formation of various features that can be observed in the solar photosphere with different sizes, lifetimes, and magnetic field strengths \citep{Cheung2014}. The most distinguished magnetic features encountered over the solar surface are sunspots \citep[see][for a full overview regarding sunspots]{Solanki2003}, which can occur and form larger entities through the solar atmosphere called active regions. It is well known that sunspots are regions harbouring a very concentrated and organised magnetic field, reaching values from $1500$~G to $3500$ G \citep{Sobotka1999_2,Solanki2003}. Thus, the magnetic field is strong enough to freeze the plasma  and is thus able to reduce the plasma dynamics and its apparent motions in the surrounding environment \citep[e.g.][]{campos2019}. The sunspot magnetic field features can be divided into two major parts, umbra and penumbra, which have different magnetic and thermodynamic properties \citep{Borrero2011b,Rezaei2012}. Because of the strong intensity contrast differences, these two components of sunspots can be easily identified by visual inspection. 

Mature sunspots are of predominant interest due to their involvement in complex magnetic field evolution and interaction with the upper atmospheric layers as well as being the sources of the most energetic solar eruptions, such as flares and coronal mass ejections \citep[e.g. reviews by][]{Green2018,Toriumi2019}. Solar pores, tiny sunspots lacking penumbrae, are worthy of attention since most of the time they constitute  the first stage of a sunspot's evolution. However, it is worth noting that not all solar pores evolve into sunspots. 

\begin{figure*}[htb]
\centering
\includegraphics[width=\hsize]{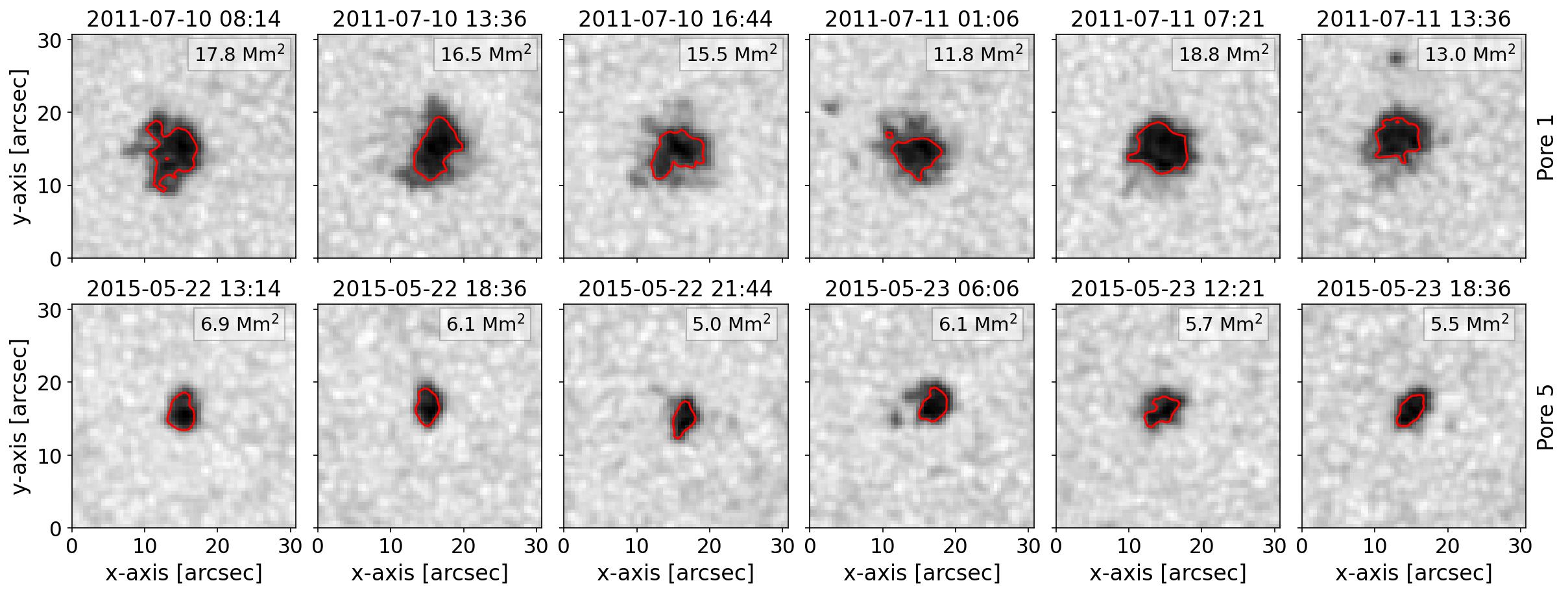}
  \caption{Time evolution of two of the studied pores with the 55\% intensity contours overplotted (red solid line). From left to right, snapshots during the pore's lifetime. The first row shows the evolution of pore 1, which revealed strong changes of the area. The second row shows pore 5, which revealed  a quasi-stable area evolution. The \href{https://drive.google.com/file/d/1YZdBgAzlzEnWCTUrdwpdZbM1bb-D18Db/view?usp=share_link}{accompanying movie} shows the evolution of each of the six pores under study.}
     \label{fig:two-pores}
\end{figure*}

It is possible that the evolution of some pores into sunspots may be attributed to the emergence of additional flux, which has the potential to alter the magnetic field topology around a pore. Such emergence events are often responsible for the creation and observation of moving magnetic features, which can be detected around certain pores prior to the formation of a penumbra. These moving magnetic features could provide a crucial link to the pre-existing magnetic topology and indicates the potential formation of filamentary substructures \citep{Keppens1996,Leka1998,Zuccarello2009,Sainz2012}.

Pores mainly harbour a simple configuration of vertical magnetic fields and are therefore particularly suitable for studying the interaction and evolution of emerging magnetic fields with a convective pattern around them \citep{Santi2022}. As a result of a convective collapse, whereby magnetic flux concentrations become too dense for convective motions to penetrate them, the trapped magnetic field can inhibit the emergence of new magnetic flux, forming the solar pores \citep{Irina2013}. Many studies go deep into the analysis of the pore's structure and dynamics while embedded in the granular pattern \citep[e.g.][and references therein]{Sobotka1999,Dorotovic2002,Vargas2010,Ermolli2017}. Recently, \citet{Gilchrist2021} probed that pores perform as magnetic waveguides, after detection of propagating magnetohydrodynamic wave activity above them. Solar pores have been the subject of numerous observational and theoretical investigations, yielding substantial advancements in our comprehension of these enigmatic features. Nonetheless, several questions related to their behaviour and characteristics persist. These include inquiries into the relationship between solar pores and sunspots, the evolution of magnetic fields in solar pores over time, the potential impact of solar pores on the upper solar atmospheric layers, and the contribution of solar pores to the overall solar activity cycle.

In this paper, we study the magnetic properties of a sample of isolated solar pores that do not evolve into sunspots using a space-based time series of photospheric filtergrams and magnetograms from the Helioseismic and Magnetic Imager \citep[HMI;][]{Hoeksema2014} on board the Solar Dynamics Observatory \citep[SDO;][]{Pesnell2012}. In Section~\ref{sec2}, we describe the data and the processing used in this study. The analysis and results are presented in Section~\ref{sec3}. Finally, we discuss our findings in the context of the magnetic properties of pores on their boundaries in Section~\ref{sec4}.

\begin{figure}[t]
\centering
\includegraphics[width=\hsize]{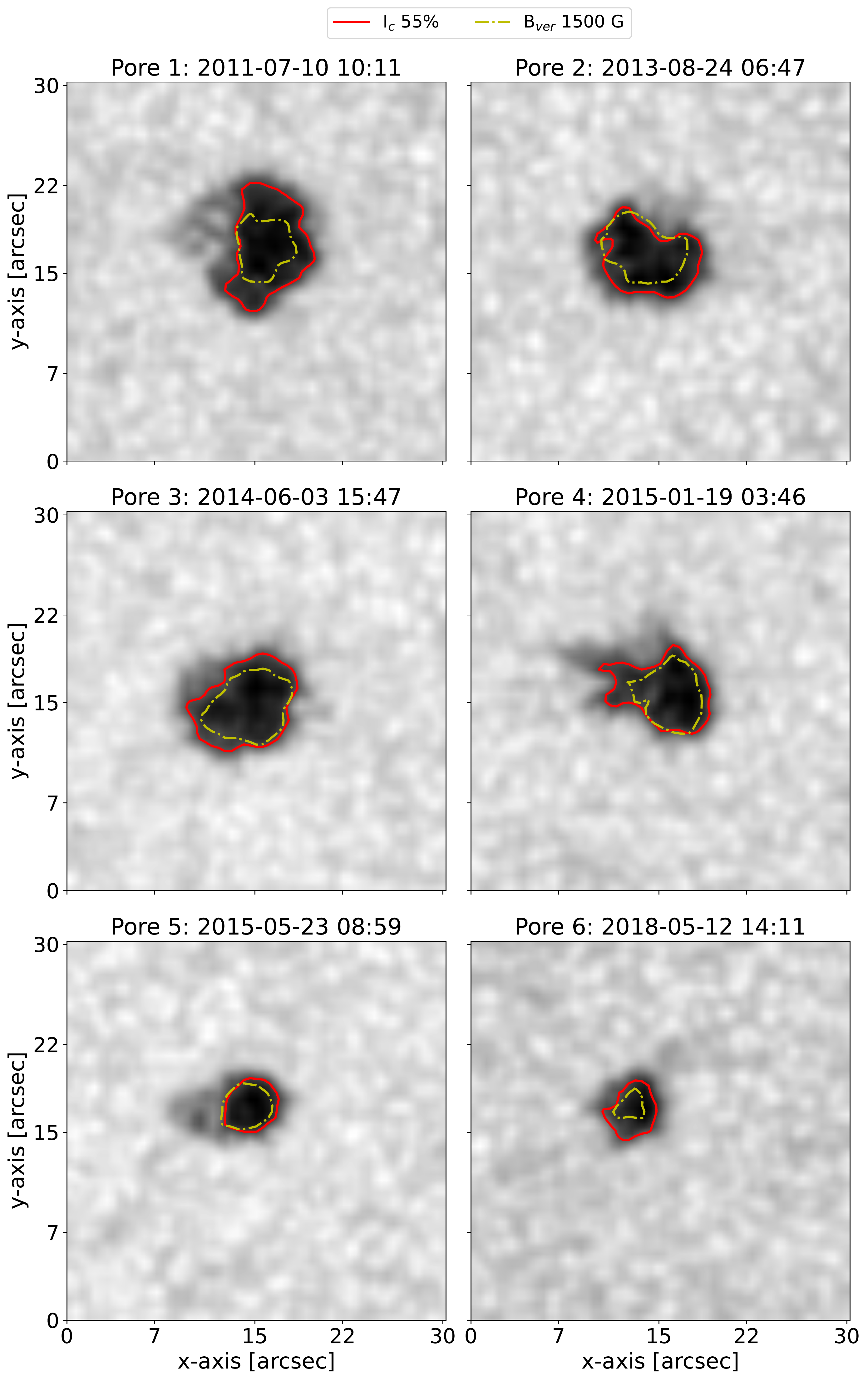}
  \caption{HMI continuum maps of the six different solar pores studied that evolved with values of $\mu$ greater than 0.9. The red contours outline an intensity value of 55\% I$_c$, and  the yellow contours outline the level of the maximum vertical magnetic field B$_{ver}$ found in the present work.}
     \label{fig:pore-sample}
\end{figure}

\section{Observational data and processing}
\label{sec2}

The task of identifying isolated solar pores exhibiting substantial areas and robust lifetimes (exceeding 24 hours) is a challenging task. To successfully execute this endeavour, we used photospheric observations obtained from a space-based telescope. We studied data from the HMI instrument on board SDO, which observes the full solar disk at $6173$ $\AA$ with a pixel resolution of $\sim0.504$ arcsec/pixel. The instrument offers a diverse range of data products, including continuum intensity images captured at cadences of 45 seconds and 720 seconds (12 minutes), line-of-sight (LOS) magnetic field measurements acquired concurrently with the intensity maps, Milne-Eddington inversions deduced from the Stokes vector observations utilising the Very Fast Inversion of the Stokes Vector \citep[][]{Borrero2011a} at a cadence of 12 minutes. The HMI data products can be accessed via the Joint Science Operations Center (JSOC) database.\footnote{\href{http://jsoc.stanford.edu/}{JSOC}}

\begin{table}[htb]
\caption{List of solar pores analysed in this work.}             
\label{tab-resum}      
\centering          
\begin{tabular}{cllc}
\hline\hline       
    Pore   & Starting    & Ending    &  \# Hours\\
    number & datetime    & datetime  &           \\
\hline                    

   1 & 2011-07-10 & 2011-07-11 &  36 \\
     & 06:59:11   & 18:58:26   &     \\
   2 & 2013-08-23 & 2013-08-25 &  36 \\
     & 23:59:07   & 11:58:22   &      \\
   3 & 2014-06-02 & 2014-06-03 &  36 \\
     & 05:59:09   & 17:59:09   &      \\
   4 & 2015-01-18 & 2015-01-20 &  35 \\
     & 17:58:54   & 04:58:09   &      \\
   5 & 2015-05-22 & 2015-05-23 &  33 \\
     & 11:59:08   & 20:47:08   &     \\
   6 & 2018-05-12 & 2018-05-13 &  27 \\
     & 05:59:09   & 08:59:05   &     \\
\hline                  
\end{tabular}
\tablefoot{Time and positional ranges of the analysed solar pores in universal time (UT).}
\end{table}
For this study, we analysed six solar pores (see Table \ref{tab-resum}) that evolved over a period of about 36 hours. We selected solar pores that evolved when they were located close to the solar disk centre and chose an observation window when the distance of the pores from the central meridian was larger than 0.9 (as given by $\mu$) in order to avoid projection effects during the analysis. Another condition for our six pores under study was that their calculated areas must be greater than $\sim$0.6 Mm$^2$ (5 pix$^2$) during the analysis time range. Moreover, another requirement was that they did not  break up into smaller micropores and spread out over larger areas during this interval. A sample of the pores is displayed in Fig.~{\ref{fig:two-pores}} and ~{\ref{fig:pore-sample}} at time instances when their evolution was close to the maximum area during their lifetimes. All the six studied solar pores were tracked over a field of view of $30\times30$ arcseconds$^2$.

The HMI continuum maps were corrected applying the centre-to-limb variation algorithm\footnote{See routine \texttt{hazel.util.i0\_allen} in \href{https://github.com/aasensio/hazel2}{HAZEL2}} to account for the limb-darkening effect and also normalised over the mean intensity value of the surrounding quiet Sun (i.e. regions where the absolute value of the LOS magnetic field is less than $50$ G). The threshold for the pore intensity boundary was defined in a manner similar to \citet{Martha}, varying the different levels for the intensity threshold in the range of 0.4-0.6 and identifying the best match to the solar pore boundary by visual inspection, which was determined as $55\%$ of the intensity values of the normalised map (example of intensity contours are shown in Fig.~\ref{fig:pore-sample} by the red solid contours). 

In an effort to compute diverse magnetic parameters on the boundary of the solar pores, we utilised the HMI vector magnetogram products denoted as "hmi.B\_720s", available on the JSOC website. In particular, we used the azimuth, inclination, and strength of the total magnetic field vector. We resolved the ambiguity of the azimuth angle through the utilisation of the disambiguation angle maps integrated into the hmi.B\_720s products. As the observations are gauged in the LOS reference system, it was imperative to re-project the azimuth and inclination magnetic angles to the local reference frame  system via the  \texttt{r\_frame\_sphduo} algorithm, a component of the AZAM software package. We subsequently calculated the vertical magnetic field component from the local reference frame inclination and total field strength. It should be noted that pores 2 and 6 (see Fig.~\ref{fig:parameters}) exhibit negative polarity, and the values for the inclination magnetic angle correspond to $180^o - \gamma$, where $\gamma$ represents the angle. Similarly, the vertical magnetic field was corrected by calculating the absolute values to the pores of negative polarity. Figure~\ref{fig:parameters} depicts the magnetic parameters employed in the present analysis, with special emphasis on the vertical magnetic field (B$_{ver}$) in order to ascertain the critical threshold value at the boundaries of solar pores. All the calculations and algorithms used in the present work have been written (or adapted) in \texttt{Python}, and the specialised library for solar physics \texttt{Sunpy}\footnote{\url{https://sunpy.org/}} \citep{Sunpy2015}.

\section{Analysis and results}
\label{sec3}
Various investigations have been conducted to establish the boundary demarcation between the umbra and penumbra \citep{Jurcak2018,Schmassmann2018}, as well as the interface between the photospheric plasma granulation and the border of solar pores \citep{Martha}. Such studies have established threshold values of the vertical magnetic field to define the boundary of these features during their lifetime evolution. In the present work, we examine the evolution of the magnetic parameters, namely, inclination, total magnetic field, and vertical magnetic field, obtained from the boundaries of the studied pores as well as the areas enclosed by the boundary contours (see Fig.~\ref{areas-all}).

\begin{figure*}[!t]
\centering
\includegraphics[width=14cm]{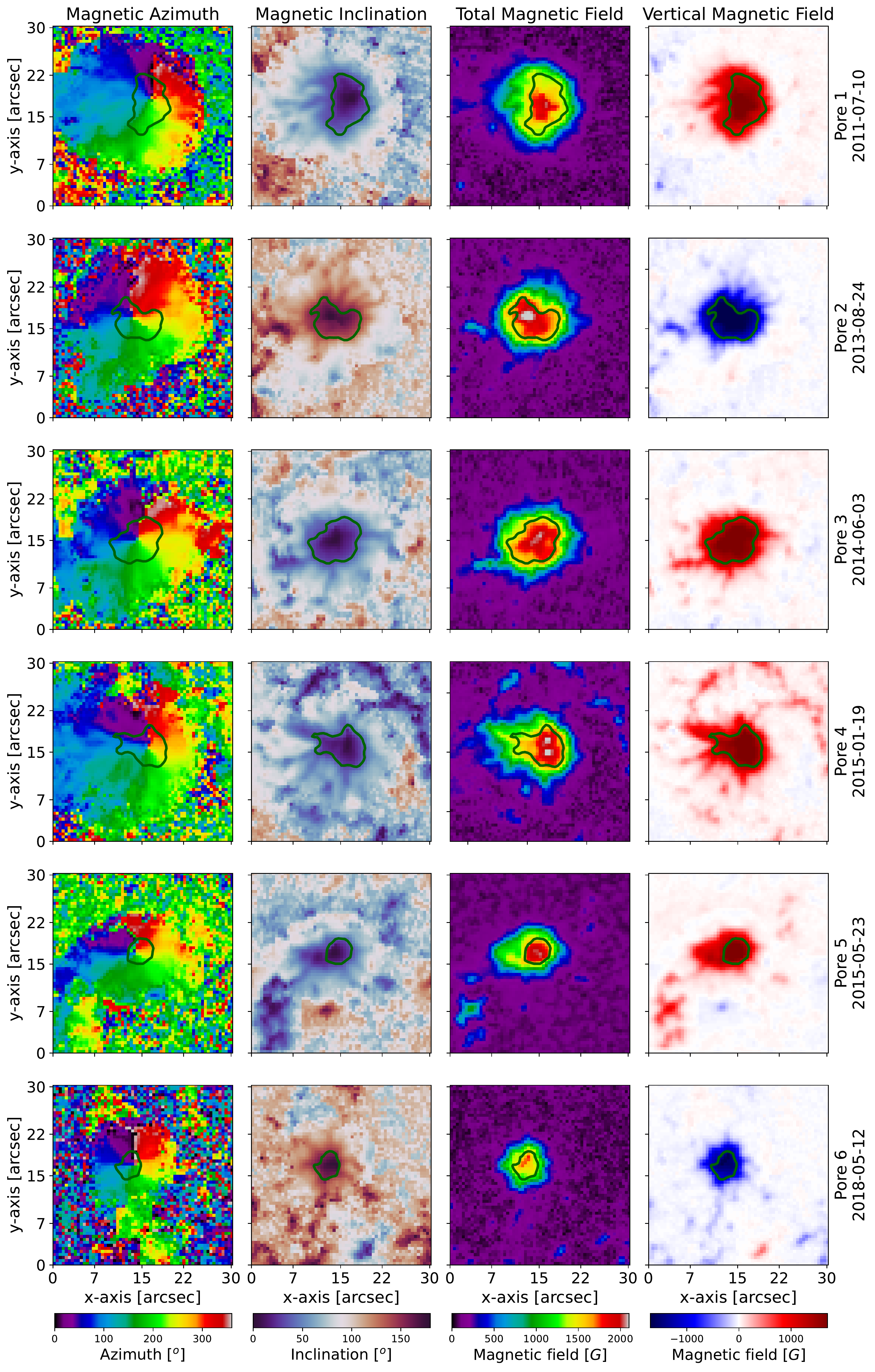}
  \caption{Snapshot of each pore during the 36-hour analysis interval. The contour in every panel outlines the pore intensity value corresponding to 55\% of $I_c$, as in Fig.~\ref{fig:pore-sample}. From left to right: azimuth, inclination, total magnetic field strength, and vertical magnetic field.
}
     \label{fig:parameters}
\end{figure*}

\begin{figure}[h]
\centering
\includegraphics[width=\hsize]{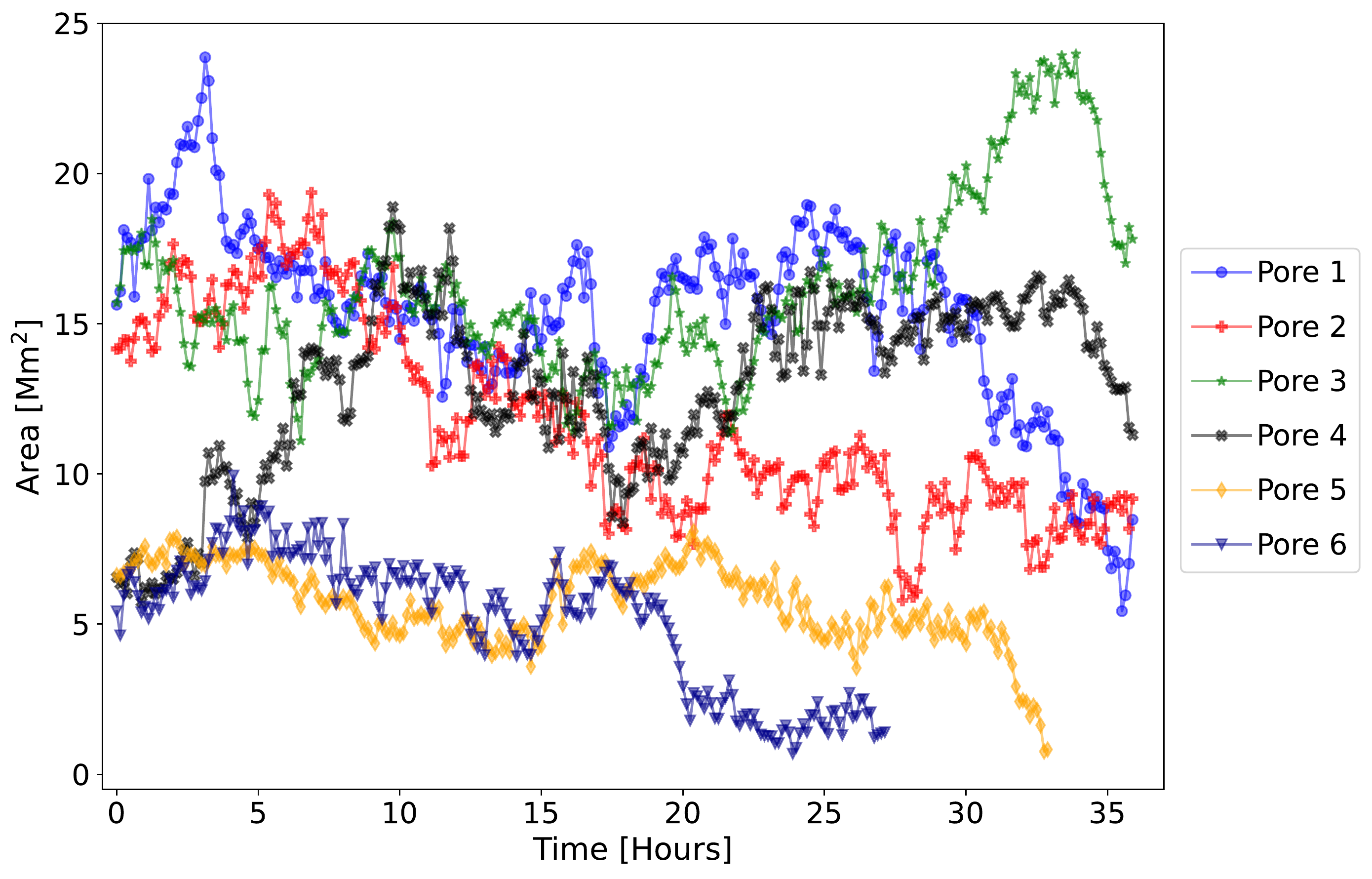}
  \caption{Temporal evolution of the areas calculated from the enclosed region determined by the contours obtained from the continuum maps for the six pores under study. (Cf. red contours in Figs.~\ref{fig:two-pores} and~\ref{fig:pore-sample}).}
     \label{areas-all}
\end{figure}

Figure 5 illustrates the temporal evolution of the analysed magnetic field parameters. These parameters were extracted and calculated as follows: First, the 55\% threshold contour was derived for the intensity maps. That contour line was then plotted over the magnetic field parameters maps (see Fig.~\ref{fig:parameters}) to extract the quantities of interest along the boundary line of the pores. In a final step, the extracted values were then averaged. The time evolution of the magnetic parameters is displayed in Fig.~\ref{all-parameters}. The evolution of the magnetic field inclination, which indicates an average of approximately $30^{\circ}$, is depicted by the green pattern in Fig.~\ref{all-parameters}.

A summary of the statistical values is presented in Table~\ref{tab-stats}. The column labelled "Average over the whole data" displays the values computed over the entire period of the boundary evolution, whereas the column labelled "Maximal vertical field" indicates the values computed when the vertical magnetic field values on the pore boundary attain the maximum value during the pore's evolution. A description of the individual magnetic parameters analysed for the six studied solar pores is presented in the following subsections.

\subsection{Magnetic field inclination $\gamma$}
 It is well established in the scientific literature that the magnetic field in solar pores is predominantly vertical \citep[e.g.][]{Rucklidge1995,Sobotka2002_1, Rempel2009,Jurcak2011}. This is also evidenced by the temporal evolution of the magnetic field inclination angle on the boundaries of solar pores, which varies on average in the range of 26-33$^\circ$, as depicted in Fig.~\ref{all-parameters}. By comparing the values in Table~\ref{tab-stats} and the areas of the pores in Fig.~\ref{areas-all}, we found that the larger the pores are, the higher is the mean inclination angle on their intensity boundaries. This is in agreement with the results obtained for sunspots (see Fig. 2 in \citet{Jurcak2018}). 

\subsection{Total magnetic field strength $B$}
For the six solar pores under study, we obtained that the values of the total magnetic field strength on the boundaries of the pores exhibit an average range of approximately 1400~G extending up to values greater than $1600$~G, with a mean standard deviation of $10$\% during their evolution. Again, larger pores tend to have greater magnetic field strengths on their boundaries, which is consistent with the behaviour found for sunspots \citep[see Fig. 2 in][]{Jurcak2018}. Figure \ref{all-parameters} provides evidence that the time evolution of these values remains relatively constant over the pore's lifetime, indicating a certain level of stability for the magnetic field values on the boundaries of the solar pores. Moreover, at the times when the vertical magnetic field attains its maximum value, the strength of the total magnetic field in the studied solar pores reaches values greater than $1600$~G, and the absolute maximum is approximately $1750$~G. This value is still smaller than the maximum values of the vertical component of the magnetic field strength found during the stable phase of the evolution of the solar pore studied by \citet{Martha} (i.e. $\sim$1920~G), who used the same intensity threshold but deconvolved HMI data. Also, \citet{Schmassmann2018} found a larger value of the total magnetic field strength on the umbral boundary of a stable sunspot: $\sim$2170~G. They used the same HMI data product as we did but employed  an intensity threshold of 50\% of the quiet Sun intensity (whereas we used a 55\% threshold). In this work, the magnetic field strength on the boundary of all the six pores during the stable phase of their evolution (when their areas did not change drastically) is thus weaker than what has been obtained in previous studies. Although the values are not directly comparable, as we either used different datasets or different intensity thresholds than previous studies, we mention them as a reference to differentiate our results.

\subsection{Vertical magnetic field strength $B_{ver}$}
The vertical magnetic field vector, $B_{ver}$, shows a behaviour similar to the total magnetic field strength. On the boundary of a pore in a stable phase, \citet{Martha} reported a maximum value of approximately 1730~G, while \citet{Schmassmann2018} reported a value of around 1630~G. We found significantly lower values for the vertical component  of the magnetic field on the boundaries of the pores under study. The maximum values are around 1500~G, and the global maximum reached in pore 3 (see Table~\ref{tab-stats}) is approximately 1570~G. However, as explained previously, one has to be cautious when comparing the values of our study with previous ones, as the methods, instruments, and data are slightly different. During the studied evolution periods of the pores, we found the mean value of the vertical component of the magnetic field to be between 1300~G and 1400~G. Similar to the evolution of the magnetic field strength, we did not observe significant variations of B$_{ver}$ during the time periods studied.

\begin{table}[htb]
\caption{Magnetic parameter statistics.}             
\label{tab-stats}      
\centering          

\resizebox{\columnwidth}{!}{%
\begin{tabular}{cllll}
\hline\hline       
    Pore   & Magnetic  &  Average over  &  Maximal \\
    number & Parameter & the whole data &  vertical field & \\
\hline\hline                    

    & Inclination [$^o$] & $29\pm6$ & $26\pm8$ \\
 1  & Total field [$G$] & $1526\pm105$ & $1655\pm125$ \\
    & Vertical field [$G$]&$ 1322\pm133$&$1480\pm150$\\
\hline
    & Inclination [$^o$]& $30\pm9$ & $23\pm5$\\
 2  & Total field [$G$]&$1610\pm138$&$1665\pm111$\\
    & Vertical field [$G$]&$1386\pm222$& $1524\pm121$  \\
\hline 
    & Inclination [$^o$]&$33\pm6$&$25\pm7$\\
 3  & Total field [$G$]&$1566\pm130$&$1602\pm73$\\
    & Vertical field [$G$]&$1312\pm163$&$1440\pm112$\\
\hline 
    & Inclination [$^o$]&$30\pm7$&$24\pm9$\\
 4  & Total field [$G$]&$1613\pm131$&$1746\pm134$\\
    & Vertical field [$G$]&$1390\pm170$&$1573\pm201$\\
\hline 
    & Inclination [$^o$]&$28\pm7$&$21\pm8$\\
 5  & Total field [$G$]&$1598\pm202$&$1653\pm210$\\
    & Vertical field [$G$]&$1394\pm216$&$1533\pm226$\\
\hline 
    & Inclination [$^o$]&$27\pm8$&$24\pm10$\\
 6  & Total field [$G$]&$1425\pm99$&$1589\pm75$\\
    & Vertical field [$G$]&$1258\pm140$&$1425\pm152$\\
\hline\hline                 
\end{tabular}
}
\tablefoot{The columns display the mean values plus or minus the standard deviation calculated from the contours obtained on the boundary of the intensity images. The inclination values are given in degrees [°], whereas the total and vertical field values are given in Gauss [G]}
\end{table}

\begin{figure*}[htb]
\centering
\includegraphics[width=\hsize]{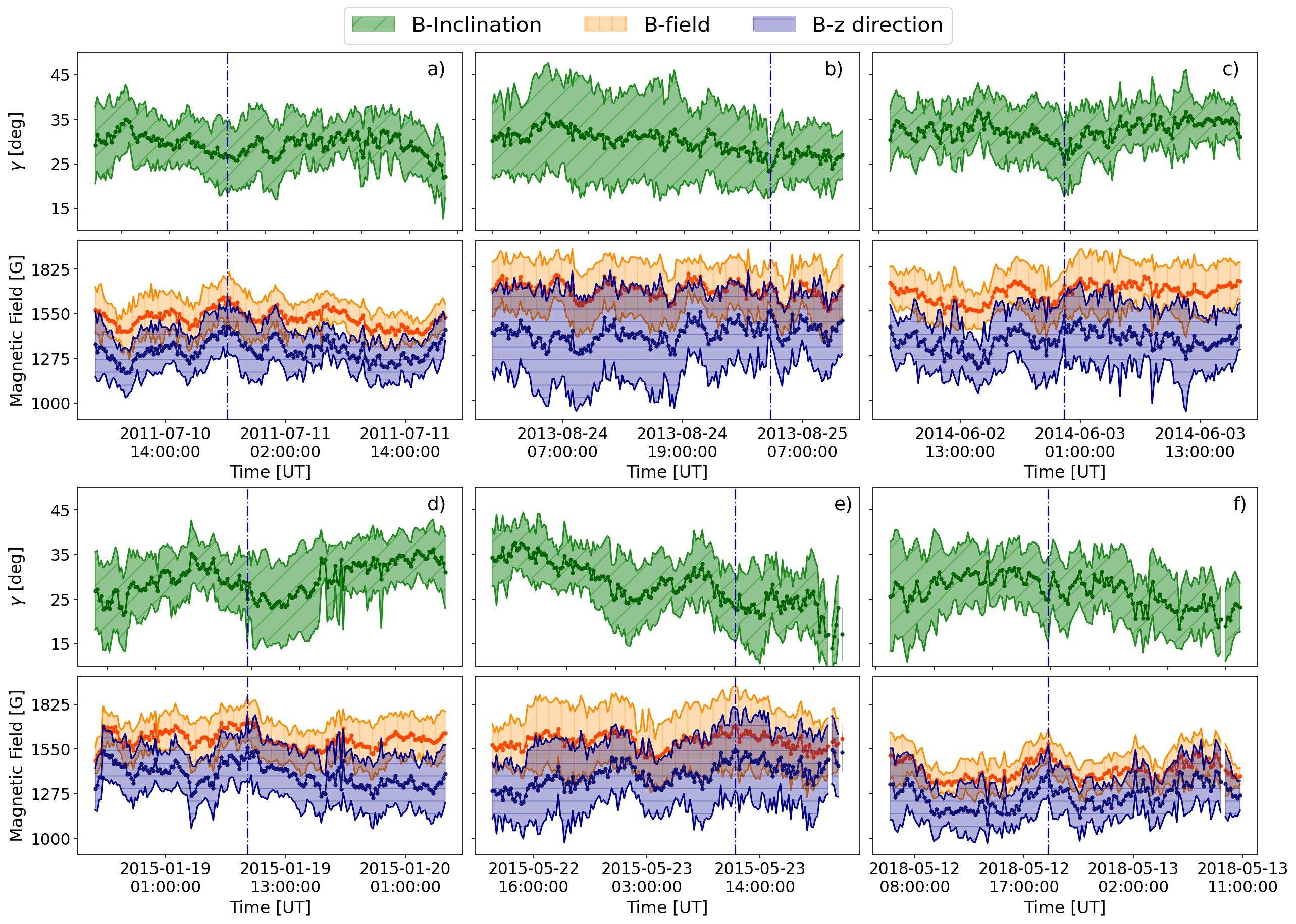}
  \caption{Time evolution of the magnetic parameters obtained from the pore's boundary contours. The main, thicker line in each plot represents the mean value obtained from the contours over each respective magnetic parameter, whereas the corresponding shaded areas refer to the standard deviation also calculated from the contours. Figure displays the magnetic field inclination angle (green pattern), the total magnetic field strength (orange pattern), and the vertical magnetic field (blue pattern) for the six pores under study.  The vertical dotted line in each panel denotes the time instant when the vertical magnetic field average over the boundary contour reached its maximum value.}  
 \label{all-parameters}
\end{figure*}

Finally, Figs.~\ref{fig:scatter} and~\ref{fig:scatter2} illustrate distinct correlations between the overall magnetic field strength, vertical field strength, magnetic field inclination, and the surface area of each pore over its lifespan. The red and blue colours correspond to two distinct time ranges defined by the moment when the vertical magnetic field reached its maximum value (indicated by the vertical dotted line in Fig.~\ref{all-parameters}) during the pore's lifetime. The straight lines displayed in the figures represent separate linear fits that were applied to the data points in each of the two distinct phases.

Overall, all analysed pores demonstrate similar behaviour, although pores 4 and 5 exhibit greater dispersions. Figure~\ref{fig:scatter} reveals that the magnetic field tends to be more vertical, with smaller inclination angles, during periods of maximum vertical magnetic field strength. In addition, Fig.~\ref{fig:scatter2} discloses the strong correlation between the magnetic inclination angle and the areas of the pores. In all cases, the maximum values of the vertical magnetic field strength are observed when the magnetic field inclination is close to either~ a global or local minimum and~coupled with a global or local maximum of the magnetic field strength. This is not surprising, as the vertical magnetic field is directly correlated with these two parameters due to it is calculated as $B_{ver}=B\cdot\cos(\gamma)$.

\section{Discussion and conclusions}
\label{sec4}
All of the six stable pores selected for this study exhibit distinct physical properties. For instance, as shown in Fig.~\ref{areas-all}, there is significant potential for variation in the temporal evolution of the pore area. Even the smallest structures in the group, pores $5$ and $6$, persist for over 24 hours before their areas decrease to values below the threshold of $\sim5$~pix$^2$ set as the minimum area. It is worth noting that pore $5$ does not exhibit a substantial change in the areas, although it does have tiny fragmentations, as determined by the intensity threshold estimation, during the analysed time period, whereas other pores, such as pores  $1$ and $2$, display significant alterations in their areas.

Despite the variability observed among the pores under study, they exhibit remarkably similar physical properties along their boundaries. Specifically, all pores examined in this study demonstrate a consistent behaviour in terms of the magnetic inclination angle evolution over the pore's lifetime, with an average value of approximately $30^o$. This result is in agreement with the statistical analysis of pores by \citet{Keppens1996}. However, we observe significantly stronger total and vertical magnetic field strengths of approximately $1560$~G and $1345$~G at the boundaries of the pores studied in this work, in comparison to the mean values reported by \citet{Keppens1996} of approximately 1400~G and 900~G, respectively. Nevertheless, the derived values for both the total and vertical magnetic field strengths are lower than those reported for stable sunspot and pore boundaries  by \citet{Schmassmann2018} and \citet{Martha}, respectively.

\begin{figure*}[htb]
\centering
\includegraphics[width=13.7cm]{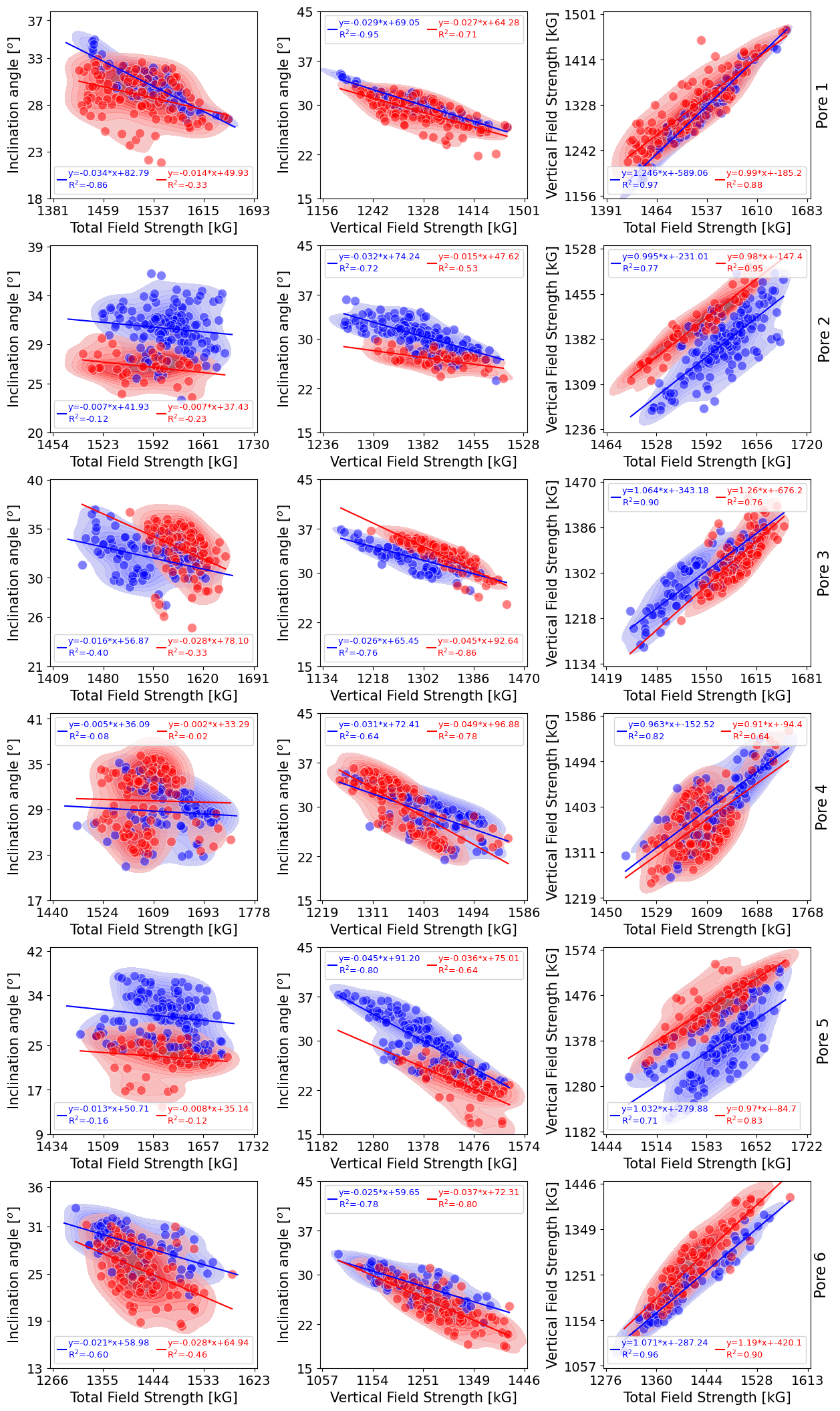}
  \caption{Scatter plot displaying the correlation between magnetic parameters derived from the boundary of the studied pores. The red colour represents the correlation between parameters during the time period before the maximum vertical magnetic field is reached (see Fig.~\ref{all-parameters}), while the blue colour represents the correlation for the time period after that instant. The shaded background contours depict the kernel density estimate plot of the correlation distribution. The insets give the parameters of the linear fit and the Pearson correlation coefficient separately for the two phases.}
     \label{fig:scatter}
\end{figure*}

\begin{figure*}[h]
\centering
\includegraphics[width=13.7cm]{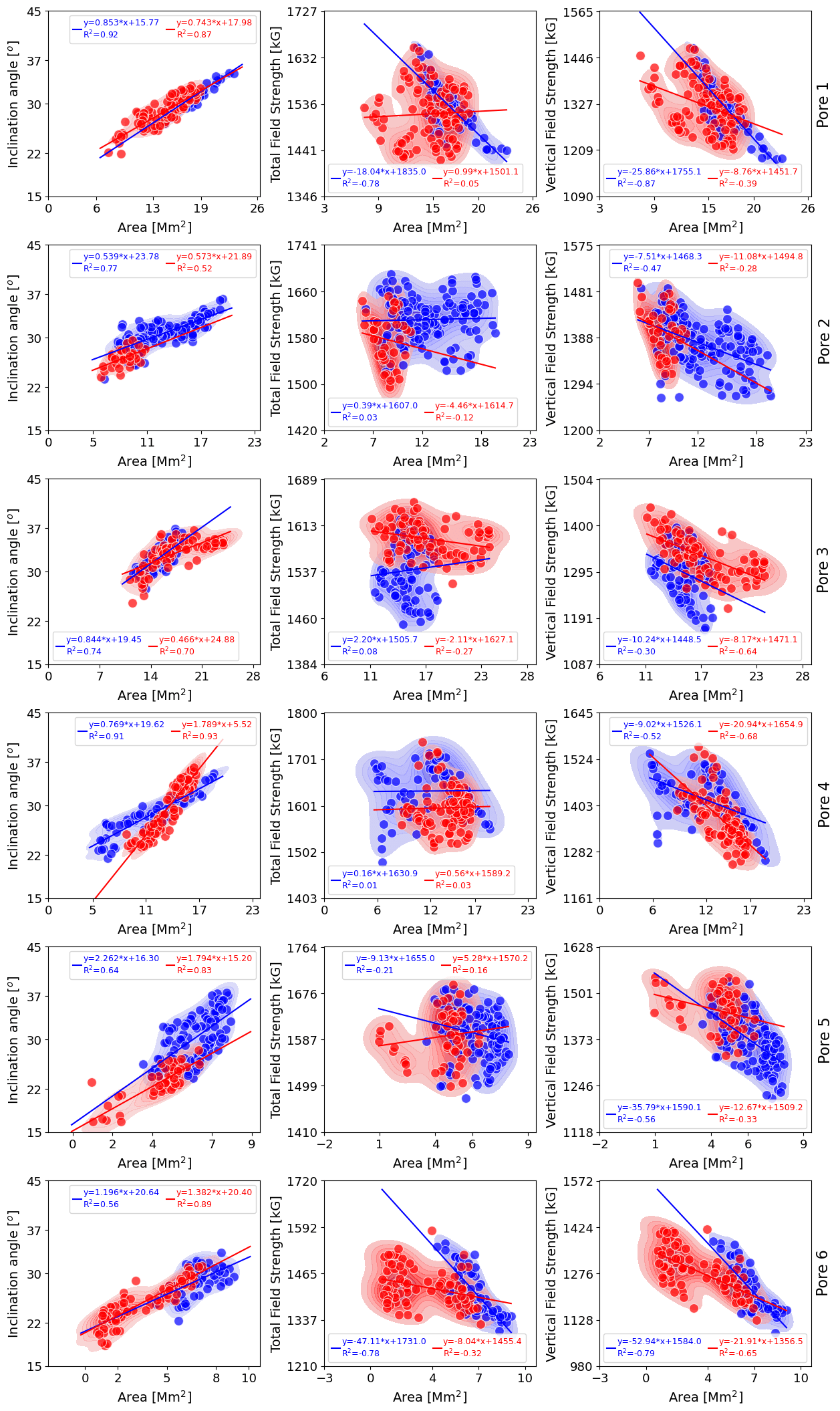}
  \caption{Scatter plot displaying the correlation between the areas calculated inside the boundary contours (see Fig.~\ref{areas-all}) and each magnetic parameter derived at the boundary of the studied pores. The colours represent the same time periods described in Fig.~\ref{fig:scatter2}. The shaded background contours depict the kernel density estimate plot of the correlation distribution. The insets give the parameters of the linear fit and the Pearson correlation coefficient separately for the two phases.}
     \label{fig:scatter2}
\end{figure*}

We note that the solar pores analysed in this work primarily consist of isolated and stable cases with weaker and more dispersed magnetic fields compared to those embedded in active regions. Active regions, which include sunspots, exhibit concentrated magnetic fields that can produce intense flares and coronal mass ejections, ultimately affecting the configuration of the entire region. Isolated stable pores tend to persist for longer periods with relatively stable magnetic field configurations and little to no accompanying activity. In contrast, active region pores are typically associated with more intense forms of solar activity and impulsive energy releases, such as flares and coronal mass ejections, and their magnetic fields are typically more dynamic and complex. Therefore, considering their differences in magnetic field strength and behaviour, isolated stable pores and active region pores can be distinguished as distinct features of the solar atmosphere. Nonetheless, it is important to note that this study only considered rather isolated stable pores, and the analysis of complex active regions exhibiting pores would be an interesting study by itself, but it is beyond the scope of this paper.

The main objective of this study was to increase the sample of the analysed boundary conditions of the magnetic field parameters in solar pores, focusing on the vertical magnetic field as the main feature. The analyses of sunspots and pores \citep{Schmassmann2018,Jurcak2018,Martha} demonstrated the vital role of the vertical magnetic field component on the stability of these structures. The range of values of the maximum vertical magnetic field obtained in this study is between $1400$~G and $1600$~G, with standard deviations between $7.8\%$ and$14.8\%$. These values are lower than those reported in previous studies, which may be due to the differences in the data type used as well as the different intensity threshold used. 

The most critical calculation when determining the boundary values is defining the contours around the structures. The spatial resolution significantly affects the results, and high spatial resolution improves the accuracy of the values obtained. The results obtained in this study differ from those of \citet{Martha}, where the vertical magnetic field values exceed the values obtained in this study by more than $100$~G. This difference could be attributed to the use of deconvolved data from SDO/HMI data \citep[see ][]{Couvidat2016}, which improves the spatial resolution of the data, and in that sense, the variation of the boundary changes considerably with respect of the original SDO/HMI maps. On the other hand, the maximum values of B$_{ver}$ found in the sample of pores studied in this work also differ. Three of the six analysed pores have a maximum B$_{ver}$ of around 1550~G (pores  2, 4, and 5), whereas the others reach lower maximum values ($\sim1448$~G). From the temporal evolution of the pore areas defined by the $55\%$ intensity threshold, it seems that only pore 5 is stable, while all the other pores undergo significant changes in their size with time. Pores 2, 4, and 5 also show the least variations in the vertical component of the magnetic field on their boundaries with time (See Fig.~\ref{all-parameters}). 

Figure \ref{fig:scatter} shows the correlation between the magnetic parameters during the two different time ranges over the lifetimes of the pores. As expected from how it is inferred, the vertical magnetic field is highly correlated with the inclination angle as well as the total strength of the magnetic field, whereby the correlation between the magnetic inclination angle and the total field shows a larger scatter.
Figure \ref{fig:scatter2} shows the correlation between the magnetic parameters and the areas obtained from the boundary contours. These plots reveal a high correlation between the evolution of the areas of the solar pores and the evolution of their magnetic inclination angles, as can be seen in the first column of Fig.~\ref{fig:scatter2}.

Defining boundary values and conditions for specific observations can enhance the development of models and initial conditions for simulations that involve the evolution of magnetic field flux tubes, such as magnetic bright points, as shown in \citet{Magyar2021}. This can provide critical inputs for comprehending the smaller scales of solar dynamics. Therefore, exploring the magnetic field properties at the boundary of solar pores, as demonstrated in this comparative study, can contribute to a deeper understanding of their dynamics.

 \begin{acknowledgements}
This research received support by the Austrian Science Fund (FWF) from the project number I 3955-N27. J. I. Campos Rozo and D. Utz acknowledge also the support from grant 21-16508J of the Grant Agency of the Czech Republic. 
All the datasets used in this work are courtesy of NASA/SDO, and they were obtained from the Joint Science Operation Center (JSOC).
 \end{acknowledgements}

\bibliographystyle{aa} 
\bibliography{bibpaper} 

\end{document}